# Large-gap insulating dimer ground state in monolayer IrTe$_2$

Jinwoong Hwang[1,2,3,#,*], Kyoo Kim[4,#], Canxun Zhang[5,6,7,#], Tiancong Zhu[5,6], Charlotte Herbig[5,6], Sooran Kim[8], Bongjae Kim[9], Yong Zhong[1,10], Mohamed Salah[1,11], Mohamed M. El-Desoky[11], Choongyu Hwang[2], Zhi-Xun Shen[3,10], Michael F. Crommie[5,6,7] & Sung-Kwan Mo[1*]

[1]Advanced Light Source, Lawrence Berkeley National Laboratory, Berkeley, CA, USA.

[2]Department of Physics, Pusan National University, Busan, South Korea.

[3]Stanford Institute for Materials and Energy Sciences, SLAC National Accelerator Laboratory, Menlo Park, CA, USA.

[4]Korea Atomic Energy Research Institute, Daejeon, South Korea.

[5]Department of Physics, University of California, Berkeley, CA, USA.

[6]Materials Sciences Division, Lawrence Berkeley National Laboratory, Berkeley, CA, USA.

[7]Kavli Energy NanoSciences Institute, University of California, Berkeley, CA, USA.

[8]Department of Physics Education, Kyungpook National University, Daegu, South Korea.

[9]Department of Physics, Kunsan National University, Gunsan, South Korea.

[10]Geballe Laboratory for Advanced Materials, Department of Physics and Applied Physics, Stanford University, Stanford, CA, USA.

[11]Physics Department, Faculty of Science, Suez University, Suez, Egypt.

[#]These authors contributed equally: Jinwoong Hwang, Kyoo Kim, Canxun Zhang

* Corresponding authors: jinwoonghwang@lbl.gov, skmo@lbl.gov

**Abstract**

**Monolayers of two-dimensional van der Waals materials exhibit novel electronic phases distinct from their bulk due to the symmetry breaking and reduced screening in the absence of the interlayer coupling. In this work, we combine angle-resolved photoemission spectroscopy and scanning tunneling microscopy/spectroscopy to demonstrate the emergence of a unique insulating 2 × 1 dimer ground state in monolayer 1$T$-IrTe$_2$ that has a large band gap in contrast to the metallic bilayer-to-bulk forms of this material. First-principles calculations reveal that phonon and charge instabilities as well as local bond formation collectively enhance and stabilize a charge-ordered ground state. Our findings provide important insights into the subtle balance of interactions having similar energy scales that occurs in the absence of strong interlayer coupling, which offers new opportunities to engineer the properties of 2D monolayers.**

**Introduction**

The layered transition metal dichalcogenides (TMDs) MX$_2$ (M = transition metal, X = S, Se, Te) provide a useful platform for studying complex electronic phases in two dimensions (2D), such as charge density wave (CDW)[1], superconductivity[2], and topological orders[3]. These electronic orders are best understood as resulting from competition between different interactions such as spin-orbit, electron-phonon, and electron-electron interactions, and so can be tuned experimentally through variation of relevant parameters[4]. In particular, by varying the thickness of TMD layers one can tune quantum confinement, screening, and even interlayer coupling, thus leading to the formation of novel ground states[5-11]. This effect is the most dramatic at the monolayer (ML) limit where interlayer coupling is completely absent, exemplified by the indirect-to-direct band gap transition in 2$H$-MoS$_2$[5,6] and 2$H$-

MoSe$_2$[7], the exotic orbital texture[8] and quantum spin liquid behavior[9] in Mott-insulating 1$T$-TaSe$_2$, and the quantum spin Hall state in 1$T$'-WTe$_2$[3,10].

Within the family of TMD materials, 1$T$-IrTe$_2$ is ideally suited for a systematic study of the hierarchy and balance of competing interactions. 5$d$ states in iridium compounds are well-known hosts of novel Mott ground states due to their strong spin-orbit coupling[12]. Bulk 1$T$-IrTe$_2$, for examples, exhibits a cascade of charge ordered states upon cooling while maintaining its metalicity[13]. Thin films of 1$T$-IrTe$_2$ have also recently been shown to exhibit a superconducting dome as a function of thickness[14,15]. Moreover, the interlayer distance in 1$T$-IrTe$_2$ is significantly shorter as compared to typical van der Waals materials[13-16], which makes it a good candidate to study the effect of the absence of interlayer coupling into the ML limit.

In this work, we report a successful molecular beam epitaxy (MBE) growth of ML IrTe$_2$ on bilayer graphene (BLG)-terminated 6$H$-SiC(0001) and characterization of its atomic and electronic structures by combined angle-resolved photoemission spectroscopy (ARPES) and scanning tunneling microscopy/spectroscopy (STM/STS). Our experimental results reveal that ML IrTe$_2$ develops a 2 × 1 dimerized atomic structure with a band gap greater than 1 eV, in stark contrast to metallic bilayer (BL) IrTe$_2$. Our first-principles calculations indicate the existence of charge and phonon instabilities in ML IrTe$_2$, suggesting that both CDW and local bond formation are responsible for the insulating dimer ground state and that non-local screening significantly impacts the band gap. Furthermore, we find that Te-Te interlayer coupling dramatically affects the phonon and charge susceptibilities in IrTe$_2$, thus playing a vital role in the metal-to-insulator transition from BL to ML.

Our findings establish ML IrTe$_2$ as a unique platform to investigate the charge order in layered 2D materials. It exemplifies a distinct ordering symmetry from that of the bulk accompanied by an opening of a full gap over the whole Fermi surface (FS). Previously studied TMD systems either have the same CDW superstructure in ML and bulk, but only ML has a full gap structure (e.g. 1$T$-TaSe$_2$)[8,9], or the ML shows a distinct CDW superstructure yet only with partial gap in a part of FS (e.g. 1$T$-VSe$_2$ & 1$T$-VTe$_2$)[17-19].

**Results**

**Epitaxial growth and structural characterization of IrTe$_2$ film**

Figures 1**a** and 1**b** show reflection high-energy electron diffraction (RHEED) images of the BLG substrate (Fig.1**a**) compared to a sub-ML coverage of IrTe$_2$ (Fig.1**b**). Clean vertical line profiles after growth indicate well-defined formation of the IrTe$_2$ film. By using the lattice constant of BLG as a reference we can estimate the lattice constant of ML IrTe$_2$ on BLG to be ~3.88 Å, which is quite comparable to the bulk value (~3.9 Å)[13,20,21]. The angle-integrated core level spectrum of our IrTe$_2$ film (Fig. 1**c**) displays sharp characteristic peaks for Ir and Te, demonstrating the film's high purity. Figure 1**d** shows a typical STM topographic image illustrating the morphology of the IrTe$_2$ films on BLG. The surface is seen to consist mostly of islands of ML IrTe$_2$, but some BL regions can also be resolved. An atomically resolved zoom-in STM topograph for ML IrTe$_2$ (Fig. 1**e**) shows a distorted 2 × 1 crystalline structure (Fig. 1**f**) that strongly deviates from the undistorted hexagonal 1$T$ phase[22]. Detailed analysis from the STM topograph (Fig. 1**e**) and its Fourier transform (Supplementary Note 1) reveal an enlarged unit cell with lattice parameters $a = 6.28 \pm 0.06$ Å, $b = 3.92 \pm 0.06$ Å, $\gamma = 85 \pm 2°$, with the definition of axes and angle as denoted in Fig. 1**e**. This distorted 2 × 1 structure is observed only in ML IrTe$_2$ and is never seen in BL IrTe$_2$

(Supplementary Figs. 3 and 5) or in bulk (Supplementary Note 2), thus suggesting a new, distinct ground state for ML IrTe$_2$.

**Electronic characterization of ML and BL IrTe$_2$**

Figure 2**a** presents the *in situ* ARPES intensity maps of ML IrTe$_2$ film taken along the M–Γ–M direction at 13 K. The ARPES band structure shows an insulating state with the valence band maximum (VBM) located at ~0.7 eV below the Fermi energy ($E_F$). The gap persists up to 300 K with no change in its magnitude (Supplementary Fig. 6). In contrast, the corresponding ARPES intensity map of BL IrTe$_2$ in Fig. 2**b** clearly exhibits a metallic state with a band crossing $E_F$ near Γ, albeit with weak intensity. The contrast between ML and BL IrTe$_2$ is further confirmed by STS d$I$/d$V$ measurements as shown in Fig. 2**c**. STS dI/dV spectra acquired on ML IrTe$_2$ exhibit a VBM ~0.67 eV below $E_F$, consistent with the ARPES results, and a conduction band minimum (CBM) ~0.35 eV above $E_F$. A statistical analysis of the STS d$I$/d$V$ spectra yields an average value for the single-particle electronic band gap of $E_g = E_{CBM} - E_{VBM} = $ ~1.02 ± 0.05 eV (Supplementary Note 4). d$I$/d$V$ spectra of BL IrTe$_2$, on the other hand, exhibit metallicity (Fig. 2**c** inset) despite weak intensity near $E_F$, also in good agreement with the suppressed ARPES intensity near $E_F$. A careful investigation of the BLG π band reveals that the influence from BLG substrate is electronically negligible to the overlaid IrTe$_2$ film (Supplementary Note 6), indicating that those properties are intrinsic in IrTe$_2$ film. However, we cannot completely rule out the structural influence from the BLG substrate as indicated from the existence of three equivalent rotational domains (Supplementary Note 4), which would require further investigation. Our combined spectroscopic measurements thus establish a large-gap metal-to-insulator transition in IrTe$_2$ when the film thickness is reduced from BL to ML, i.e., as the Te-Te interlayer coupling is eliminated.

**ARPES spectra and calculated band structure of ML IrTe$_2$**

We further characterized the unexpected insulating ground state of ML IrTe$_2$ using polarization dependent ARPES measurements. Figures 3**a,d** display the ML IrTe$_2$ ARPES intensity maps measured along the M–Γ–M direction using *s*- and *p*-polarized photons, respectively. The second derivatives of the spectra are also shown in order to more clearly visualize weak spectral features (Figs. 3**b,e**). Several "X-shaped" band features, which come from averaging of domains (Supplementary Note 7), are observed around the Γ-point that are best resolved in the second derivative maps. The polarization-dependent maps show strong intensity contrast arising from the orbital character of the measured electronic states. For example, the ARPES intensity distribution with *s*- (*p*-) polarization is more intense from 2.5 to 4.0 eV (0.8 to 2.5 eV) below $E_F$. This indicates that in-plane Ir orbital states (e.g., $d_{x^2-y^2}$ and $d_{xy}$ that are more pronounced with *s*-polarization[4,8,10] due to the rotated domains) lie deeper within the valence bands (Supplementary Note 11). For comparison, the corresponding polarization-dependent BL IrTe$_2$ spectra are shown in Supplementary Fig. 4. Not only these spectra show clear metallic behavior, but much simpler band dispersion than ML one, which also verifies the structural integrity among layers in the BL films.

To better understand the origin of the insulating ground state observed in ML IrTe$_2$, we carried out density functional theory (DFT) calculations. By starting from the 2 × 1 supercell of undistorted ML 1*T* structure and relaxing a-b lattices as well as the atomic positions based on the identified 2 × 1 structure from the STM topograph (Fig. 1**e** and Supplementary Fig. 1), we obtained a final distorted structure in which both the Ir atoms and the top/bottom Te atoms are fully dimerized (Fig. 1**f**). The resulting lattice parameters ($a$ = 6.39 Å, $b$ = 4.01 Å, γ = 83°) reasonably match the values obtained from the STM topograph (Fig. 1**e** and Supplementary Fig. 1), and so we identify the experimentally observed 2 × 1 structure with the dimer state found in the calculations. The ARPES measurements were then simulated by calculating the Ir orbital-projected band structure for the relaxed dimer state. To compare with experiment the band structure was unfolded into 1$T$ ML Brillouin zone and averaged over three different rotational domains with equal weight (Figs. 3**c,f**) (see Supplementary Note 7). Reasonable agreement is found between the DFT calculations and the ARPES maps (Fig. 3) including both the X-shaped band structure features and the polarization-dependent intensity, thus confirming the 2 × 1 dimerized structure of Fig. 4**b**.

**Discussion**

Although the ARPES intensity maps of ML IrTe$_2$ are quite well described by our DFT calculation, the estimated DFT band gap of 0.55 eV is significantly smaller than the experimental value of 1.02 ± 0.05 eV (see Supplementary Fig. 13). Since both Ir and Te atoms have strong spin-orbit coupling (SOC), the SOC may play a vital role in the electronic property, as has been the case for iridium oxides where the strong SOC leads to a novel Mott insulating phases[12,23] and potential superconductivity[24]. However, including SOC does not rectify this discrepancy (Supplementary Fig. 11) since the effect of SOC is suppressed by the direct Ir-Ir dimerization which lifts the three-fold $t_{2g}$ orbital degeneracy of Ir[25]. Including an additional on-site Coulomb interaction $U$ in the calculation also does not resolve the situation since this surprisingly decreases the band gap for this system and even leads to a metallic ground state (Supplementary Fig. 12). This unconventional response to the addition of an on-site $U$ can be understood in terms of strong Ir-Te hybridization, which leads to delocalized electronic states that invalidate a Hubbard-type approach[26-28]. We find that the enhanced ML band gap is best accounted for by the GW$_0$ approach which takes into account the long-range screening effects that arise from the extended orbital nature and reduced dimensionality of this system[27,28]. Figure 4c shows a comparison of the electronic band structures calculated by the DFT and $GW_0$ approximations. The DFT band dispersions are not qualitatively modified in the $GW_0$ calculation, but the band gap is significantly enhanced by strong self-energy corrections arising from reduced screening in 2D and strong *p-d* hybridization[27,28] (see Supplementary Fig. 14). The $GW_0$ calculation, in fact, overestimates the band gap by 0.7 eV (the $GW_0$ gap is ~1.7 eV compared to the experimental gap of $1.02 \pm 0.05$ eV; see Supplementary Fig. 13), but this could be due to additional screening from the BLG substrate[11] that is not taken into account in the calculation.

We now discuss the driving mechanism of the 2 × 1 Ir dimerization in ML $IrTe_2$. In order to provide theoretical insights, we performed first-principles calculations of the phonon dispersion and the electronic susceptibility for ML $1T$-$IrTe_2$, a natural hypothetical high temperature unit cell of $IrTe_2$. A clear sign of a phonon softening is observed at the M-point of the calculated phonon dispersion (Fig. 4d), and the electronic susceptibility (Fig. 4e and Supplementary Fig. 10d) shows a dominant peak at the M-point, indicative of a Fermi surface nesting CDW with $\vec{q}_{CDW} = \vec{M}$[1,29] (Supplementary Fig. 10g). However, such nesting-type weak coupling CDW alone usually only causes small partial gap[1,29], so it would not fully account for the heavily reconstructed electronic structure with the large full gap observed in both ARPES and STS (Fig. 2), contrary to the small partial gap in weak coupling CDW cases[1,29]. Moreover, the contraction of the Ir-Ir distance in dimers is strikingly ~20% shorter than non-dimerized $1T$-$IrTe_2$ (3.12 Å vs. 3.88 Å) (Figs. 4a,b), which is much larger distortion compared to a conventional CDW 1~7%[1,29].

The additional factor here that leads to this "oversized" dimerization is the local bond formation in the Ir dimers. Generally, Ir compounds with partially filled $t_{2g}$ orbitals with edge-shared structure prefer locally forming a direct Ir-Ir singlet due to their extended 5$d$ orbitals[25,30,31]. The ML 1$T$-IrTe$_2$ shows the edge-sharing octahedra structure (Fig. 4**a**) and only Ir$^{4+}$ (5$d^5$) valence state (Fig. 1**c** inset and Supplementary Fig. 4**e**) with one hole in $t_{2g}$ states, due to the complete absence of the Te-Te interlayer coupling. Once the perturbations in charge and lattice channels are triggered by the nesting instability from Te character (Supplementary Fig. 10**a**), hexagonal 1$T$-IrTe$_2$ is spontaneously transformed to 2 × 1 structure in the ML, unlike stable hexagonal 1$T$-IrTe$_2$ structure in bulk[13,22,32]. The structural distortion amplifies wavefunction overlap between Ir atoms, facilitating the formation of covalent-type Ir-Ir dimerization[25]. This chemical bonding mechanism further stabilizes the total energy of the ML IrTe$_2$ by energy gain from dimerization[25]. As a result, there is an abrupt change in electronic structure such as the unusual large band gap from bonding-antibonding splitting (Supplementary Fig. 14) observed in our spectroscopic results. Such a mechanism is supported by the ARPES data which exhibits strongly bounded in-plane Ir orbitals states (Fig. 3). The quenching of dimerization in 1$T$-IrTe$_2$ with increased thickness is also nicely explained by our theoretical picture. Adding additional layers to the ML induces Te-Te coupling between layers that has dual effects of altering the Ir valence state and distorting the IrTe$_2$ Fermi surface (see Supplementary Figs. 4**e** and 10**e**). This results in the break of Ir singlet and the lift of nesting conditions (see Supplementary Notes 8 and Note 10) to stabilize the metallic state in BL IrTe$_2$.

In conclusion, our combined ARPES, STM/STS, and first-principles study of ML 1$T$-IrTe$_2$ has revealed a 2 × 1 dimer structure in the ML with a band gap larger than 1eV, which establishes it as a unique platform to investigate the charge order in layered 2D materials. Our theoretical calculations suggest that ultra-strong dimerization arises from positive feedback between nesting-type CDW and local bond formation in ML IrTe$_2$. We show that strong Te-Te interlayer coupling plays a vital role in the insulator-to-metal transition from ML to BL IrTe$_2$ by strongly affecting both the Ir valence state and the Fermi surface nesting properties. Our findings provide a compelling case for the emergence of a novel IrTe$_2$ ground state caused by elimination of strong interlayer coupling in the ML limit. This creates new possibilities for the discovery and control of novel electronic phases in 2D van der Waals materials and their heterostructures.

**Methods**

**Thin film growth and in-situ ARPES measurement**

The ML and BL IrTe$_2$ films were grown by molecular beam epitaxy (MBE) on epitaxial bilayer graphene on 6*H*-SiC(0001) and transferred directly into the ARPES analysis chamber for the measurement at the HERS endstation of Beamline 10.0.1, Advanced Light Source, Lawrence Berkeley National Laboratory. The base pressure of the MBE chamber was $3 \times 10^{-10}$ torr. High-purity Ir (99.9%) and Te (99.999%) were evaporated from an e-beam evaporator and a standard Knudsen effusion cell, respectively. The flux ratio was Ir:Te = 1:30, and the substrate temperature was held at 380 ˚C during the growth. This yields the growth rate of 1 hour per monolayer monitored by *in situ* RHEED. After growth, the IrTe$_2$ film was annealed at 390 ˚C for 2 hours to improve the film quality. ARPES data was taken using a Scienta R4000 analyzer at base pressure $3 \times 10^{-11}$ torr. The photon energy was set at 63 eV for *s*-polarization and 70 eV for *p*-polarization with energy and angular resolution of 18-25 meV and 0.1˚, respectively. The spot size of the photon beam on the sample was ~100 μm ×100 μm. To achieve high quality ARPES data of ML IrTe$_2$, we grew a low coverage film (less than 0.5 ML) and performed in-situ ARPES measurement. Potential charging effect from the insulating samples has been monitored by checking reference spectra with varying photon flux.

**STM/STS measurements**

To protect the IrTe$_2$ film from an exposure to air during the transfer to the STM chamber, we sequentially deposited Te and Se capping layers with a thickness of ~100 nm on the film before taking the samples out of the ultrahigh-vacuum (UHV) system of Beamline 10.0.1. After transferal of the sample through air to the STM UHV chambers, the samples were annealed at 300 ˚C for 2 hours in UHV system to remove the capping layer before STM measurement. All STM/STS measurements were performed in a commercial Omicron LT-STM held at $T = 4.7$ K. STM tips were prepared on an Au(111) surface and calibrated against the Au(111) Shockley surface state before each set of measurements to avoid tip artifacts. d$I$/d$V$ spectra were recorded using standard lock-in techniques with a small bias modulation $V_{RMS} = 2$–20 mV at 613 Hz. All STM images were edited using WSxM software[33].

**Density Functional Theory Calculations**

For the structural optimization and phonon calculations we used the projector augmented wave (PAW) method as implemented in the Vienna ab initio simulation package (VASP)[34,35] within the PBEsol-GGA exchange correlation functional[36]. GW$_0$ calculations are performed on top of DFT results by wannierizing relevant bands with 100 frequency points and 128 virtual orbitals were used. Ultra-soft PAW potentials with appendix(_GW) in vasp.5.2 were used. We performed the phonon calculations including spin-orbit coupling with the supercell approach implemented in the Phonopy package[37]. The dynamical matrix has been obtained with $2 \times 2 \times 1$ supercell for ML and BL IrTe$_2$ as well as an $8 \times 1 \times 1$ supercell for the BL using the Hellmann-Feynman force theorem. For the analysis of band structure, orbital characters, charge susceptibility, and unfolded spectra, we additionally employed the Full Potential Local orbital (FPLO) package[38].

**Data availability**

The data that support the plots within this paper and other findings of this study are available from the corresponding authors upon reasonable request.

**Acknowledgements**

C. Z., T. Z. and C. H. thank Xuehao Wu for technical support. The work performed at the Stanford Institute for Materials and Energy Sciences and Stanford University (thin film characterization) was supported by the Office of Basic Energy Sciences, the US Department of Energy under Contract No. DE-AC02-76SF00515. The work performed at the Advanced Light Source (sample growth and ARPES) was supported by the Office of Basic Energy Sciences, the US Department of Energy under Contract No. DE-AC0205CH11231. The work performed at UC Berkeley (STM/STS measurements) was supported as part of the Center for Novel Pathways to Quantum Coherence in Materials, an Energy Frontier Research Center funded by the US Department of Energy, Office of Science, Basic Energy Sciences. C.G. H. and J. H. acknowledge fellowship support from the National Research Foundation of Korea (NRF) grant funded by the Korea government (MSIP) (Grant No. 2021R1A2C1004266 and 2020K1A3A7A09080369). M.S. was funded by a scholarship (JS3809) from the Ministry of Higher Education of Egypt. K. K. acknowledges support from Internal R&D program at KAERI (Grant No. 524460-22) (susceptibility calculations) and NRF (Grant No. 2016R1D1A1B02008461) (band structure analysis). S. K. acknowledges support from NRF (Grant No. 2019R1F1A1052026) and KISTI Supercomputing Center (Project No. KSC-2020-CRE-0255) (phonon calculations). B. K. acknowledges support from NRF (Grant No. 2018R1D1A1A02086051) and KISTI Supercomputing Center (Project No. KSC-2019-CRE-0231) (GW calculations).

**Author contributions**

J.W.H. and S.-K.M. initiated and conceived the research. J.W.H. performed the thin film growth and ARPES measurements with the help from Y.Z., M.S. and M.E.-D and under the supervision of S.-K.M., Z.-X.S., and C.G.H.. J.W.H analyzed the ARPES data. C.Z., T.Z., C.H. carried out STM/STS measurements and analyses under the supervision of M.F.C.. K.K., S.K., B.K. performed DFT calculations and theoretical analyses. J.W.H., K.K., C.Z., M.F.C, and S.-K.M wrote the manuscript with the help from all authors. All authors contributed to the scientific discussion.

**Competing interests**

The authors declare no competing interests.

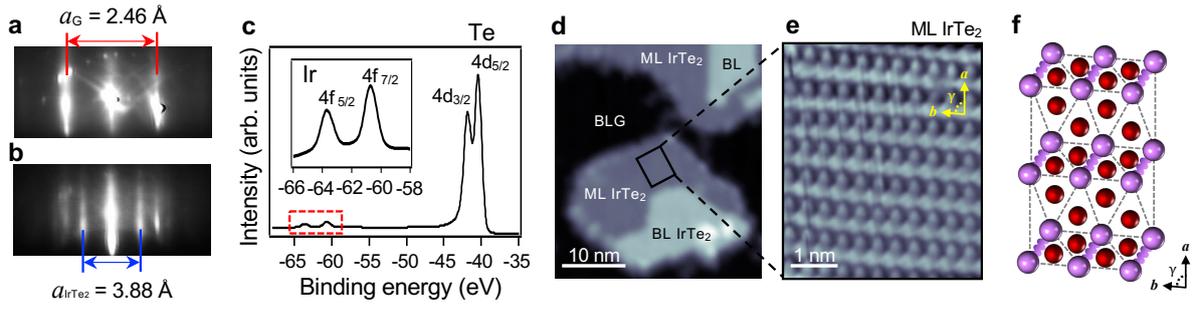

**Figure 1 Characterization of the epitaxial grown ML IrTe$_2$. a,b**, RHEED images of **a**, BLG substrate and **b**, sub-ML IrTe$_2$. **c**, Core level spectra of ML IrTe$_2$ measured at 13 K using 110 eV photons. The inset is a close-up for the range marked by the red dashed box. **d**, Typical STM topographic image of IrTe$_2$ on BLG substrate ($V_s$ = 1.5 V, $I_0$ = 0.01 nA, $T$ = 4.7 K). **e**, Atomically-resolved STM image of ML IrTe$_2$ ($V_s$ = 1 V, $I_0$ = 0.25 nA, $T$ = 4.7 K). **f**, Schematics of a top view of the distorted crystal structure of ML IrTe$_2$. Purple and red balls represent Ir and Te atoms, respectively. Purple wavy lines represent dimerized Ir atoms.

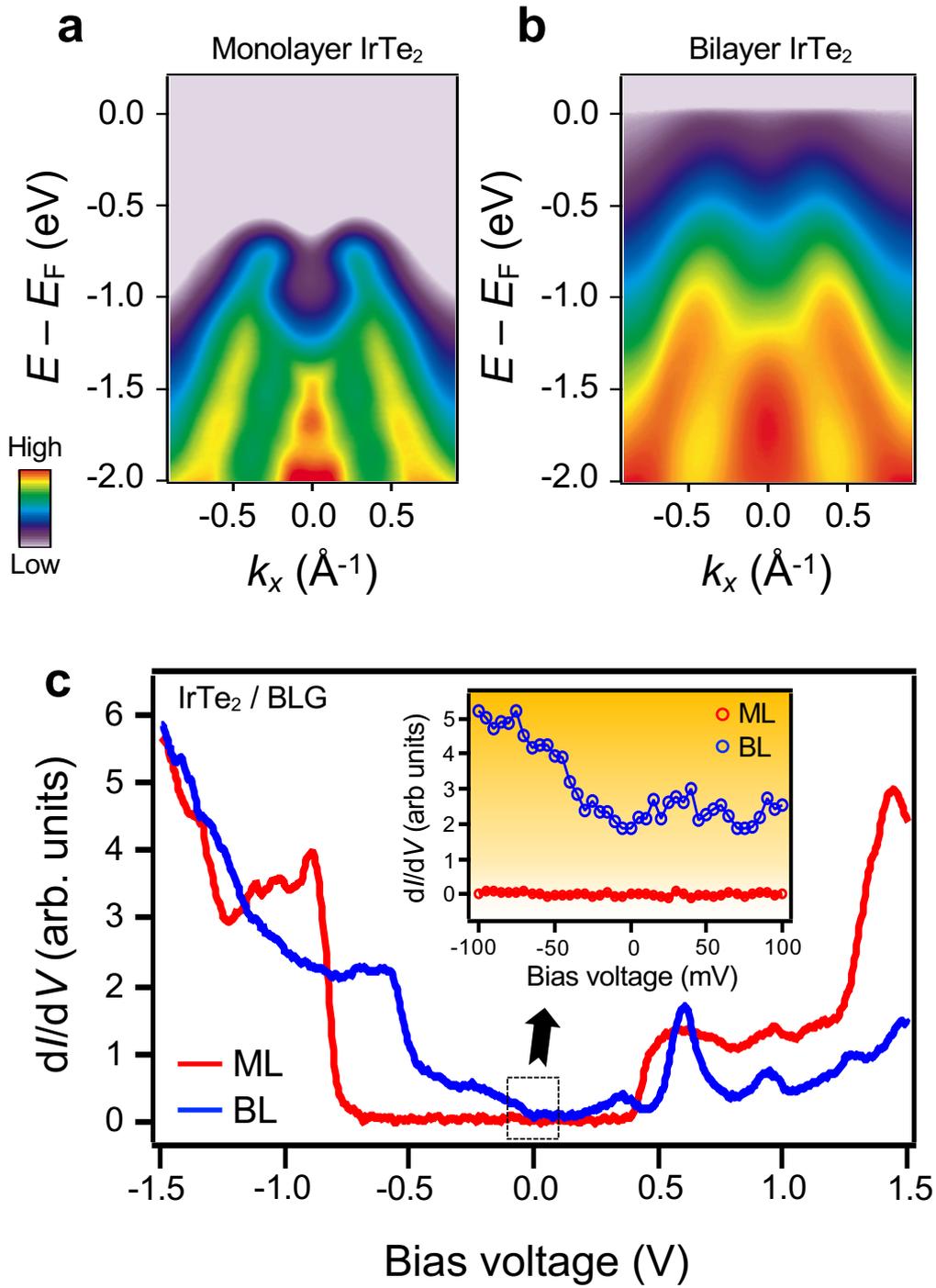

**Figure 2 Thickness-induced metal-to-insulator transition in IrTe$_2$.** **a,b**, ARPES intensity maps of **a**, ML, and **b**, that of BL IrTe$_2$ taken along the M–Γ–M direction using *p*-polarized photons (*T* = 13 K). **c**, The STS d*I*/d*V* spectra for ML and BL IrTe$_2$ ($V_s$ = 1.5 V, $I_0$ = 0.01 nA, modulation voltage $V_{rms}$ = 10 mV, *T* = 4.7 K). The inset is a close-up look of the black dashed box near $E_F$.

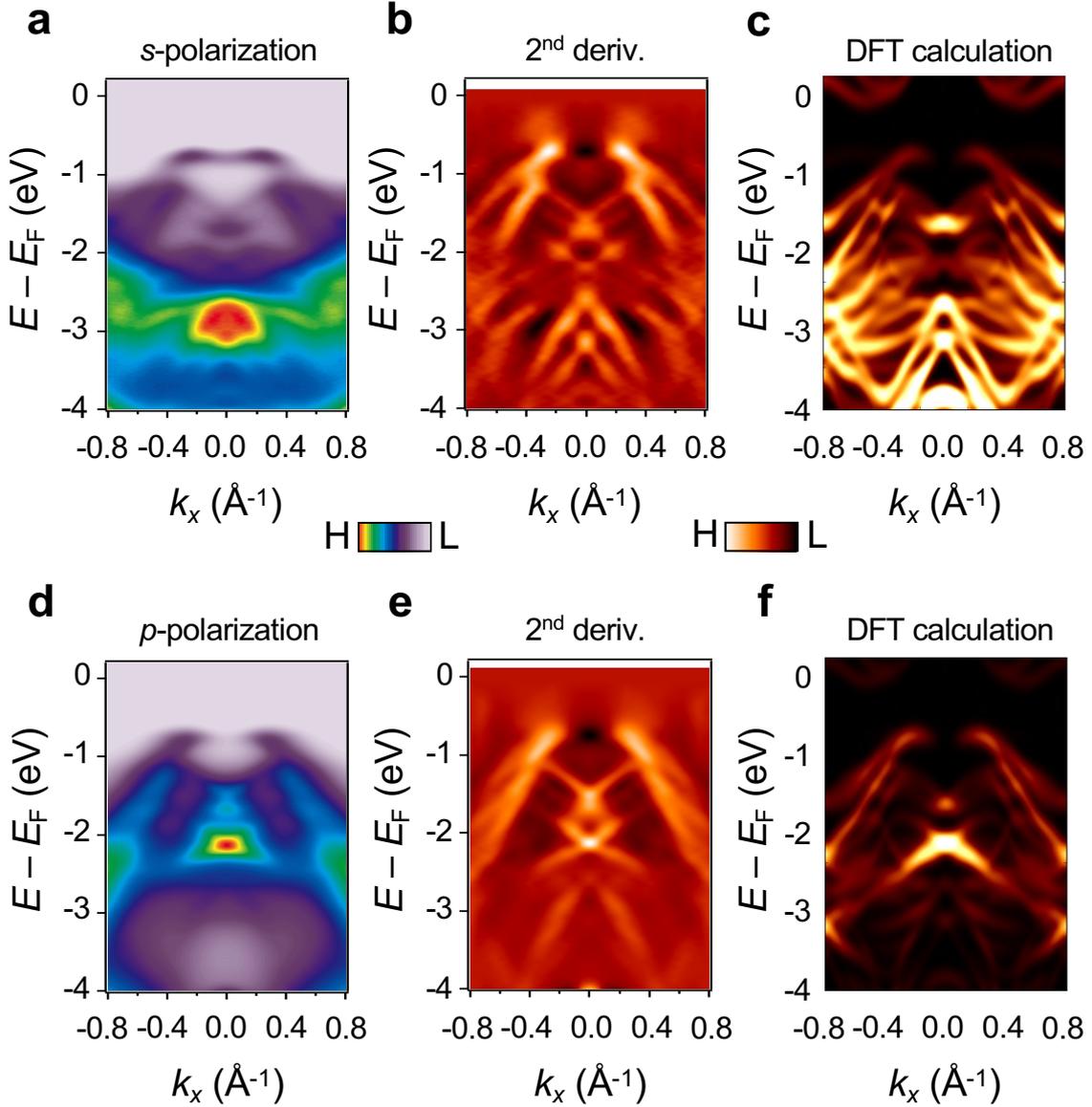

**Figure 3 Polarization dependent ARPES and electronic structure of ML IrTe$_2$**. **a**, ARPES intensity map of ML IrTe$_2$ taken along the M–Γ–M direction using *s*-polarized photons ($T$ = 13 K). **b**, Its second derivative with respect to momentum. **c**, In-plane Ir orbitals ($t_{2g}$, $d_{x^2-y^2}$)-projected DFT band structure. **d**, ARPES intensity map of ML IrTe$_2$ taken along the M–Γ–M direction using *p*-polarized photons. **e**, Its second derivative with respect to momentum. **f**, Out-of-plane Ir orbital ($d_{z^2}$)-projected DFT band structure.

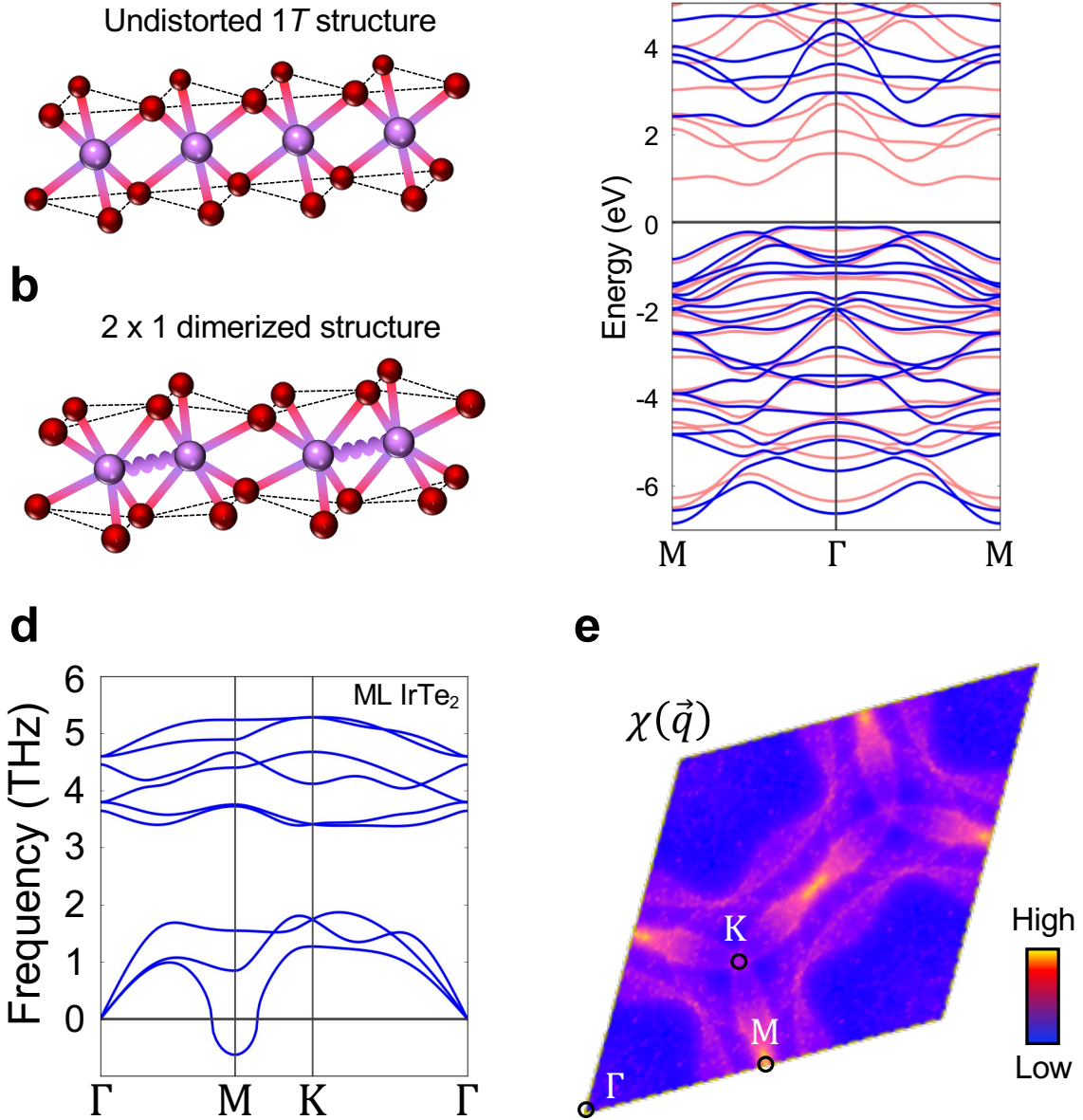

**Figure 4 Origin of the large-gap 2 × 1 dimerized structure**. **a,b**, Schematics of the crystal structure of **a**, ML undistorted 1$T$-IrTe$_2$ and **b**, 2 × 1 Ir dimerized ML IrTe$_2$. Purple wavy line represents a Ir dimerization. **c**, Comparison between calculated DFT (pink) and GW$_0$ (blue) band structures. **d**, Calculated phonon spectrum of ML 1$T$-IrTe$_2$ along its high symmetry directions. **e**, Real part of the electronic susceptibility $\chi(\vec{q})$ of ML 1$T$-IrTe$_2$.